\documentclass[10pt,conference]{IEEEtran}

\usepackage{mdframed}
\usepackage{float}
\usepackage{stackengine}
\usepackage{xcolor}
\usepackage{balance}
\definecolor{lightestgrey}{rgb}{0.9,0.9,0.9}
\usepackage{graphicx}
\usepackage{enumitem}
\usepackage[hidelinks]{hyperref}
\usepackage{slantsc}
\usepackage{listings}
\newcommand{\thegap}[0]{0.2\baselineskip}
\mdfsetup{%
  innertopmargin=4pt,
  innerbottommargin=4pt,
  innerleftmargin=4pt,
  innerrightmargin=4pt,
}

\usepackage{booktabs}
\newcommand{\tablestyle}[0]{\sffamily\centering}

\newenvironment{finding}[0]{\begin{mdframed}\noindent\textbf{Finding:}}{\end{mdframed}}
\newenvironment{survey}{\begin{mdframed}\noindent}{\end{mdframed}}

\usepackage{xparse}
\NewDocumentEnvironment{implementation}{o}{
  \vspace{\thegap}\noindent\begin{minipage}{\linewidth}\ttfamily\noindent
}{
    \IfNoValueF{#1}{\hfill{}{\sffamily\upshape\footnotesize (scored #1)}}
  \end{minipage}
}
\NewDocumentEnvironment{whyquotation}{o}{
  \noindent
  \begin{minipage}{\linewidth}\centering
  \vspace{\thegap}
  \begin{minipage}{\dimexpr \linewidth - 2em}\itshape\raggedright
}{
    \IfNoValueF{#1}{~{\sffamily\upshape\footnotesize (scored~#1)}}
  \end{minipage}
  
  \vspace{\thegap}
  \end{minipage}
}
\newcommand{\etal}[0]{et~al{.}}
\newcommand{\hltext}[1]{\underline{#1}}
\newcommand{\examplewhy}[2]{\addlinespace\multicolumn{5}{p{\dimexpr \linewidth-5\tabcolsep}}{\colorbox{lightestgrey}{\begin{whyquotation}[#1]\itshape {``#2''}\end{whyquotation}}}}

\newcommand\blfootnote[1]{%
  \begingroup
  \renewcommand\thefootnote{}\footnote{#1}%
  \addtocounter{footnote}{-1}%
  \endgroup
}

\title{``Do this! Do that!, And nothing will happen''\\Do specifications lead to securely stored passwords?}

\author{\IEEEauthorblockA{Joseph Hallett, Nikhil Patnaik, Benjamin Shreeve and Awais Rashid}
  \IEEEauthorblockN{University of Bristol}}

\begin{document}
\maketitle

\begin{abstract} 
Does the act of writing a specification (how the code should behave) for a piece
of security sensitive code lead to developers producing more secure code? We
asked 138 developers to write a snippet of code to store a password: Half of
them were asked to write down a specification of how the code should behave
before writing the program, the other half were asked to write the code but
without being prompted to write a specification first.  We find that explicitly
prompting developers to write a specification has a small positive effect on the
security of password storage approaches implemented. However, developers often
fail to store passwords securely, despite claiming to be confident and
knowledgeable in their approaches, and despite considering an appropriate range
of threats. We find a need for developer-centered usable mechanisms for telling developers how to store passwords: lists of what they \emph{must} do are not working.
\end{abstract}

\section{Introduction}

\blfootnote{The quote in the title is attributed
to Harry S. Truman~\cite{neustadt1960presidential}, and is used as the
opening quote to chapter 6 in \emph{The Mythical
Man-Month}~\cite{brooks1972mythical}.}
Developers struggle to store passwords securely.  Naiakshina~\etal{}
have repeatedly shown that developers do not build in security unless
explicitly asked to do so (and even then typically do so
poorly)~\cite{naiakshina-2017-password,naiakshina-2018-deception,naiakshina-2019-if-you-want}.
In organizations, one can support developers coding securely through
code review, and acceptance testing---but not all developers work in
teams and many work alone on their own
projects~\cite{meyer-2017-characterizing,vanlinden2020schrodinger}.

Developers continue to seek guidance on how to handle passwords.  In a survey of
developers' posts on Stack Overflow (a popular developer question and answer
site) Barua~\etal{} found that posts related to authentication and security
(including password storage) and were one of the top 20 topics on the site
and accounted for 2.1\% of all questions~\cite{barua2014developers}.
Furthermore in a survey of just security-focussed posts on Stack Overflow,
Yang~\etal{} found that the most viewed of all security-focussed posts related
to passwords~\cite{yang2016security}, with each post viewed on average 2,731
times.  Whilst there are alternatives to passwords~\cite{hardt2012oauth}, many
developers still appear to be working with passwords and implementing password
storage in their apps and software. As well as working with passwords they are
seeking guidance on how to do it \emph{right}.

It is well established that writing a specification before implementing it leads
to code that is of a higher
quality~\cite{boehm1984verifying,parnas1972technique,spolsky2004joel,demarco1979structure,brooks1972mythical,meyer-1992-design-by-contract}.
Since specification writing is beneficial for \emph{quality}, does the act of
writing one also improve developers' security practices? Naiakshina's work
suggests that developers only consider security aspects if explicitly
prompted~\cite{naiakshina-2019-if-you-want}; but if we try to continuously
prompt developers to \emph{work securely} we risk security
fatigue~\cite{furnell2009recognising,parkin2016fatigue}. Since specification is
an established developer practice, this paper seeks to explore whether the act
of writing any form of specification primes developers to program securely: in other words whether giving developers time to make a plan (however formally or informally) leads them to either recalling more about how to store passwords, or to recall that a standard exists and to check.

To test this we recruited 138 developers from  an online platform for recruiting
participants for studies, and asked them to write code to store a password in
whatever language they were most comfortable with.  Half the developers were
asked to write down a specification---any form of specification from a formal
definition~\cite{lamsweerde2000formal} to a prose
description~\cite{padegs1981system}---of how they would implement this before they
were allowed to write their solution, the rest were allowed to write their
implementation immediately.  We scored the security of their implementations
using Naiakshina's end-user password storage
criteria~\cite{naiakshina-2017-password} (Figure~\ref{fig:naiakshina}), which is
itself based on NIST~SP~800-63-3~\cite{nist-800-63-3}, and analyzed their
written justifications of their choices and threats considered.

Specifically, we address the following research questions:
\begin{description}
\item[RQ1] Does specification writing lead to a measurable
improvement in password storage method?
\item[RQ2] What approaches do developers take when implementing
password storage and what do they typically remember and forget?
\item[RQ3] How do developers justify their implementation approach and
what threats do they consider?
\end{description}

\noindent Our key findings are as follows:
\begin{itemize}[leftmargin=*]
\item Developers who were explicitly prompted to write a
specification, stored their passwords \emph{slightly} more securely
than those who were not prompted ($p=0.027$, $r_{rb}=0.209$).
\item Only 38\% of developers remembered to hash passwords, 14\% remembered
to salt them, but other secure password storage practice was largely
absent (Figure~\ref{fig:naiakshina}).
\item Developers think they are storing passwords correctly, but their
scores according to Naiakshina's criteria (Figure~\ref{fig:naiakshina}) do not indicate best practice.
\item If given time to reflect, some developers do realize that there are threats to stored passwords and that their solutions may not be secure.
\end{itemize}
\noindent
Our novel insights are in examining whether specification is a useful tool for priming for security related tasks, and how developers justify the code they write and the threats they consider with respect to the specifications they write. Analyzing the developer's rationale suggests that whilst some developers consider appropriate threats, their knowledge of best practice is out-of-date and that current cryptographic guidelines~\cite{nist-800-63-3}.  Providing lists of what developers must do is not working. Instead we must fit the task to the developer and provide usable mechanisms for password storage.

\begin{figure} \centering
  \begin{mdframed}\raggedright
    \begin{itemize}[leftmargin=*]
      \item The end-user password is salted (+1) and hashed (+1).
      \item The derived length of the hash is at least 160 bits long (+1).
      \item The iteration count for key stretching is at least 1,000 (+0.5) or 10,000(+1) for PBKDF2 and at least $2^{10}$ for bcrypt (+1).
      \item A memory-hard hashing function is used (+1).
      \item The salt value is generated randomly (+1).
      \item The salt is at least 32 bits in length (+1).
    \end{itemize}
    \end{mdframed}
  \caption{Naiakshina's end-user password storage assessment
criteria~\cite{naiakshina-2017-password}, copied verbatim.  A score
$\mathbf{\geq 6}$ indicates industrial best practice.}
  \label{fig:naiakshina}
\end{figure}

\section{Background and Related Work}

\subsection{Benefits of specification}

The benefits of program specification are well established in both the
academic and engineering communities.  Spolsky notes their benefit
saying:
\begin{quote} ``If you don’t have a spec, you will always spend
more time and create lower quality code."~\cite{spolsky2004joel}
\end{quote}
Brooks~Jr{.} also notes the benefit of a
specification:
\begin{quote} ``Careful function definition, careful
specification, and the disciplined exorcism of frills of function and
flights of technique all reduce the number of system bugs that have to
be found."~\cite{brooks1972mythical}
\end{quote} Dromey suggested that quality models and requirements
specifications could lead to an improvement in software
quality~\cite{dromey1996cornering}.  Haigh and Landwehr have suggested
that by building code to security specifications (drawing analogy to
US \emph{building codes}) we can reduce the vulnerability in software
systems~\cite{haigh2015building,landwehr2017building}.
Polikarpova~\etal{} found that twice as many bugs were found when code
was written with a \emph{strong
specification}~\cite{polikarpova-2013-what-good}.  Mohanani~\etal{},
however, found that specifications can lead to developers blindly
following them without considering why the rules
exist~\cite{mohanani2014requirements}.
Our work seeks to demonstrate that the act of writing a specification creates an implicit priming effect that can impact a developer's approach to security.

\subsection{Work on password security}

There is a large body of work surrounding passwords, but a small
subset that addresses how developers perform password storage and
present analysis of the process. Password storage is a feature
generally supported by cryptographic libraries. The usable security
community has studied the developers' interaction with the
cryptographic APIs.

Naiakshina~\etal{} ran the first qualitative usability study to observe how 20
computer science students address the task of password
storage~\cite{naiakshina-2017-password}. They concluded that
participants consider functionality before security. Unless
participants are primed, they do not think the task of password
storage requires a secure solution. On the other hand, participants
who were primed to consider security used various hash functions and
different algorithms to secure their password. For the participants
who were primed, none of their solutions met the academic standards of
the time. Cryptographic frameworks offer password storage as an opt-in
feature. This means the developers needs to understand cryptography to store passwords. 10\% of the non-primed
participants attempted a secure solution for password storage while
70\% of the primed participants attempted a secure solution.
On asking the non-primed
students about the security oversight, they replied that they would
have implemented secure storage if they were writing code for a
commercial product. To address this insight Naiakshina~\etal{} conducted a field study
with freelance developers. Like students, freelance developers do not
consider security for password storage, unless prompted. Both students
and freelance developers have misconceptions about secure password
storage, however interestingly freelance developers show a wider range
of these misconceptions. Freelance developers often stored passwords
with Base64, confusing encoding functions with hash functions, a
misconception they shared with end-users. Naiakshina~\etal{} conclude, that
even when developers believe they are coding for companies they seldom
store the password securely without
prompting~\cite{naiakshina-2019-if-you-want}.  
Acar~\etal{} conducted an experiment with GitHub developers to establish if they
are an accurate representation of developers in general for security-based
developer studies. The GitHub developers were asked to perform password storage
securely. The solutions included the storage of plain-text passwords, use of
static salts, use of unsafe hashing algorithms~\cite{acar2017security}.  Our
work goes beyond Acar~\etal{}'s and Naiakshina~\etal's work by examining
developer's rationale for their password storage implementations and finds that,
whilst developers aren't storing passwords securely, they \emph{think} they're
following best practices.

Oesch~et~al{.} evaluated 13 popular password managers and their
solutions for handling the 3 main stages of a password's life-cycle;
password generation, storage, and auto-fill. Their evaluation of
password storage showed that developers stored information in plain-text,
left metadata unencrypted, and used insecure
defaults~\cite{oesch2020then}. Our work compliments this by diving deeper into \emph{why} developers do not engage with best practice.
There is a large body of work on 
end-user passwords and their security~\cite{shay2010encountering, ur2012helping, ur2015added,
ur2016users, ur2017design, bonneau2012science, egelman2013does,
komanduri2011passwords, weir2010testing}. In contrast our work focuses on developer's approaches to storing passwords.

\subsection{Work on secure programming}

Weir~\etal{} looked at the prevalence of security assurance techniques (including threat assessment and code review) among Android developers~\cite{weir2020needs}.  They found that between only 22--30\% of Android developers used these techniques despite a high perceived need for security.  We found that $\sim$56\% of developers claimed to write a specification without prompting.
Fischer~\etal{} examined the amount of code copied from Stackoverflow, and its security~\cite{fischer2017stack}. They found that 15\% of Android apps contained vulnerable code copied from Stackoverflow.  We found that $\sim$8\% of developers copied from Stackoverflow specifically, but that a further $\sim$12\% copied from other online sources.

Many vulnerabilities arise due to developers misusing cryptographic
libraries.
Nadi~\etal{} performed an empirical investigation into challenges
developers face when using Java cryptographic APIs. Based on the
analysis of 100 Stack Overflow posts, 100 GitHub repositories and a
survey of 48 developers, they found that developers find cryptographic
features such as encryption and digital signatures difficult to
program. they also found that APIs are generally perceived to be too
low-level for developers~\cite{nadi-2016-jumping-through-hoops}.

Egele~\etal{} studied the integration of cryptographic APIs in Android
applications. They found errors in 88\% of the applications. CryptoLint was
introduced as a static analysis tool to find these
errors~\cite{egele-2013-empirical}.  Patnaik~\etal{} performed a thematic
analysis of 2491 Stack Overflow posts from developers seeking help with using 7
cryptographic libraries, and found 16 usability
issues~\cite{patnaik2019usability} that could be related to Green and Smith's
earlier work that proposes usability principles for cryptographic APIs.  show
that developers find cryptographic APIs challenging to use.  We find that as
well as struggling with APIs developers are not clear on what they need to do to
store passwords securely, following current guidelines~\cite{nist-800-63-3}.

\section{Method}

We used a between-subjects design to explore whether the act of
specification writing results in more secure code being produced.

\subsection{Study Design}

To test the effect specification writing had on implementation we designed a
study where developers would implement the part of an app's code for storing
passwords. We chose password storage as a task as it is security relevant,
implementable within a relatively short space of time and is a common task with
plenty of guidance available that most developers would have encountered in
their work.

Our study was implemented as a set of online tasks and questions (to capture
rationale).  Developers were randomly assigned a grouping (either
\emph{specification} or \emph{no-specification}) and shown the following
scenario:

\begin{survey}You are working on the backend of an application.
Users create an account on the app, and login before being allowed to
use the program.
  The application is complete bar one task: writing the login system
users use to authenticate with the app.  You have been tasked with
implementing this part of the app.
You decide to start with storing the users' passwords. Your boss
trusts your judgment when it comes to implementing this feature.
\end{survey}
Developers in the \emph{specification} group were then asked to write
a specification for how the password should be stored.

\begin{survey}You decide to start by writing a specification for
how the password should be stored, and to note down any special
requirements and implementation details.  You are provided with a
username and password, and they have been checked to see that they are
valid text.

  Describe your specification below.
  You can describe your specification using formal notation, informal
notes, a list, mathematical notation or any other method.  If you draw
a picture as part of your specification, please say so and say what is
shown.
\end{survey}
Both groups were then asked to implement the password storage using
whichever language they wished.  If they used a real programming
language they were asked to note it.

\begin{survey}You start writing the password storage method.  You
have been given the password the user wishes to use and you need to
store it so that it can be checked whenever users try and login.
  You are given a username and password.  Both have been checked to be
valid text (i.e. neither empty nor containing bad characters)
  Write code (or pseudocode) to implement the password storage.  Your
code doesn't need to be compilable or syntactically correct but should
illustrate your general approach to the implementation.
\end{survey}
Developers in the \emph{no-specification} group were then asked if
they had made some form of specification or plan before starting their
implementation (without being asked to).  Those that indicated that
they did, were asked to describe their specification and their results
were added to those of the \emph{specification} group.
Both groups were then asked to provide a rational for their coding
approach in a free text box. They were asked if they considered
\emph{what threats might attack a stored password}, and, whether they
\emph{referred to any standards for password storage} when
implementing the code.  Finally, participants were asked whether they
had any formal qualifications in software engineering, or computer
science; and, to rate their knowledge of security and cryptography on
a 5-point Likert scale,
and briefly describe their security and development experience.

\subsection{Analysis}

To analyze the data we scored each of their implementations using Naiakshina's
metric~\cite{naiakshina-2017-password} and compared the average score between
different groups using the Mann Whitney $U$ test (a rank-based non-parametric
test to explore if two groups are distinct~\cite{mann1947test}) to test for
significance and to calculate the effect sizes (using the rank-biserial
correlation~\cite{cureton1956rank}).  To analyze developer's rationale and
threat models we asked developers to describe them and analyzed them
qualitatively using a grounded theory approach~\cite{Strauss1998,Glaser1967}.

\subsection{Recruitment and Ethics}

Developers were recruited from \emph{Prolific Academic} and were
screened, by Prolific, based on their familiarity with computer
programming.  Developers were offered a financial reward for
completing the study of $\pounds${5}, inline
with the \emph{living wage} in our country. All developers who
completed the study were paid for their work.

Ethical approval for the study was sought from and granted by
Bristol University.  No personal data was
collected, and demographic data was deleted after coding and
validation. Data is available by request.

\subsection{Limitations and Threats to Validity}

We acknowledge the following limitations and threats to our study:

\begin{itemize}[leftmargin=*]
\item Our developers were recruited by \emph{Prolific Academic} and as such, may not be representative of how developers as a
whole behave. Other studies have also used similar populations for studying passwords and developers~\cite{kelley2012guess,ur2016users,naiakshina-2020-conducting-security-developer-studies}.
\item Developers may not know how to store a password, and may not be
aware that it is a security related task.  We mitigate this by
qualitatively analyzing the developers' rationale behind their code.
\item Developers who were not prompted to write a specification,
may opt to write a specification anyway.  To correct for this we asked developers not in the specification group if they
 wrote a specification, after their implementation.  We assume that the
specification produced by the unprompted group is similar to the
prompted group (and we ask them to describe it), but this may not be
the case and some participants may retrospectively write a
specification.
\item We ask the developers about their qualifications and experience,
however all data is self-reported and may not be accurate.
\item We asked developers to implement password storage and 99
developers (72\%) did so.  19 developers (14\%) instead appeared to
write code implementing password authentication (how one would
check if a password was correct) but from which their approach to
password storage could be seen. A further 15 developers (11\%) stored
the password, but did so only checking if the password contained a
suitable range of letters, numbers and symbols, 3 (3\%) approached the
problem by retransmitting their passwords over HTTP\footnote{Three
appeared to have copied the question from:
\url{https://stackoverflow.com/questions/19999417/password-storage-in-code-how-to-make-it-safe}.},
and 1 insisted the passwords be stored \emph{alphabetically}. We include all in our analysis, as they were all conceivably ways a developer may approach storing passwords.

\item Scoring implementations according to Naiakshina's criteria could
introduce subjectivity. To mitigate this, one author scored and then another author independently rescored all the
implementations and calculated Cohen's Kappa (a measure of inter-rater
reliability~\cite{landis1977measurement}). The kappa-value indicates
\emph{almost perfect agreement}
($\kappa=0.94$)~\cite{landis1977measurement}.
Similarly, our codebooks, whilst grounded in data, were likely influenced by the coder's background and experiences.  Using our
codebooks a separate coder independently re-coded the entire dataset.  We found
\emph{substantial agreement} ($\kappa=0.72$)
with our coding for developers' explanations for their implementations
(Table~\ref{tab:whycodebook}) and \emph{almost perfect agreement}
($\kappa=0.84$) with our coding for the
threats developers considered.
\item We measure developers' password storage approaches using Naiakshina's
  criteria, but this poses a construct validity threat.  We chose this metric as
  it has been used in prior
  work~\cite{naiakshina-2017-password,naiakshina-2019-if-you-want}, and on a
  NIST standard for password storage~\cite{nist-800-63-3}.   We mitigate this
  threat by qualitatively analyzing \emph{why} developers wrote the code they
  did as well as their implementations.
\end{itemize}

\section{Quantitative Results}

\begin{table*}\tablestyle
  \caption{Distribution of scores for password storage methods by
different groups.  Absolute values are given in (parentheses). The
{specification} group consists of two-subgroups: those that we
explicitly prompted for a specification, and those that we did not
prompt but reported writing one unprompted.  A score of 6 or more is
considered to be following best practice.}
\begin{tabular}{l r r | l l l l l l l l | l l l}
  \toprule
  \multicolumn{1}{c}{Group} & \multicolumn{1}{c}{Count} & \multicolumn{1}{c}{$>0$} & \multicolumn{1}{c}{0} & \multicolumn{1}{c}{1} & \multicolumn{1}{c}{2} & \multicolumn{1}{c}{3} & \multicolumn{1}{c}{4} & \multicolumn{1}{c}{5} & \multicolumn{1}{c}{6} & \multicolumn{1}{c}{7} & \multicolumn{1}{c}{$\mu$} & \multicolumn{1}{c}{$\sigma$} \\
  \midrule
  Specification & 104 & 43 & 59\% (61) & 20\% (21) & 11\% (11) & 5\% (5) & 4\% (4) & 0 & 2\% (2) & 0 & 0.83 & 1.30 \\
  \hspace{1em}Prompted specification & 61 & 27 & 56\% (34) & 15\% (9) & 15\% (9) & 7\% (4) & 5\% (3) & 0 & 3\% (2) & 0 & 1.03 & 1.51 \\
  \hspace{1em}Unprompted specification & 43 & 16 & 63\% (27) & 28\% (12) & 5\% (2) & 2\% (1) & 2\% (1) & 0 & 0 & 0 & 0.53 & 0.88 \\
  Unprompted & 77 & 26 & 66\% (51) & 25\% (19) & 6\% (5) & 1\% (1) & 1\% (1) & 0 & 0 & 0 & 0.47 & 0.79 \\
\hspace{1em}No Specification & 34 & 10 & 71\% (24) & 21\% (7) & 9\% (3) & 0 & 0 & 0 & 0 & 0 & 0.38 & 0.65 \\
  \addlinespace
  Used standard & 36 & 16 & 56\% (20) & 19\% (7) & 11\% (4) & 8\% (3) & 3\% (1) & 0 & 3\% (1) & 0 & 0.94 & 1.41 \\
  No standard & 102 & 37 & 64\% (65) & 21\% (21) & 10\% (10) & 2\% (2) & 3\% (3) & 0 & 1\% (1) & 0 & 0.64 & 1.10 \\
  \addlinespace
  Formally qualified & 59 & 27 & 54\% (32) & 22\% (13) & 15\% (9) & 5\% (3) & 3\% (2) & 0 & 0 & 0 & 0.81 & 1.09 \\
  Not formally qualified & 79 & 26 & 67\% (53) & 19\% (15) & 6\% (5) & 3\% (2) & 3\% (2) & 0 & 3\% (2) & 0 & 0.65 & 1.26 \\
  \addlinespace
  Overall & 138 & 53 & 62\% (85) & 20\% (28) & 10\% (14) & 4\% (5) & 3\% (4) & 0 & 1\% (2) & 0 & 0.72 & 1.19 \\
  \bottomrule
\end{tabular}

  \label{tab:scores}
\end{table*}

\begin{table}\tablestyle \setlength{\tabcolsep}{4pt}
  \caption{Comparison between groups using the Mann-Whitney \emph{U}
test.}  \begin{tabular}{l l | r r c}
\toprule
\multicolumn{1}{c}{Group 1} & \multicolumn{1}{c}{Group 2} & \multicolumn{1}{c}{$U$} & \multicolumn{1}{c}{$p$} & \multicolumn{1}{c}{$r_{rb}$} \\
\midrule
Prompted Specification & Unprompted & 1948 & 0.024 & 0.171 \\
\addlinespace
Specification (all) & No specification & 1495 & 0.061 & 0.154 \\
\hspace{1em}Prompted specification & No specification & 820 & 0.027 & 0.209 \\
\hspace{1em}Unprompted specification & No specification & 675 & 0.247 & 0.077 \\
\addlinespace
Used standard & No standard & 1638 & 0.135 & 0.108 \\
\addlinespace
Formal qualification & No qualification & 2018 & 0.061 & 0.134 \\
\bottomrule
\end{tabular}

  \label{tab:mwu}
\end{table}

\begin{table}\tablestyle
  \caption{Frequency different points in Naiakshina's criteria were
observed compared to the whole population.  No answer scored a
half-point for key-stretching. (Absolute values).}  \begin{tabular}{lrr}
\toprule
\multicolumn{1}{c}{Criteria} & \multicolumn{2}{c}{Observations} \\
\midrule
Hashed & 38\% & (53)\\
Salted & 14\% & (19)\\
Hash length & 7\% & (9)\\
Key stretching & 2\% & (3)\\
Memory-hard hashing & 1\% & (1)\\
Random salt & 7\% & (10)\\
Salt length & 3\% & (4)\\
\bottomrule
\end{tabular}

  \label{tab:metric}
\end{table}

\begin{table}\tablestyle \setlength{\tabcolsep}{4pt}
  \caption{Co-occurrences of points in Naiakshina's criteria (i.e. 36\%
of all participants who hashed their password also salted their
passwords).  (Absolute values).}
\newcommand{\vertically}[1]{\rotatebox{90}{#1}}
\begin{tabular}{l|ccccccc}
\multicolumn{1}{c}{} & \vertically{Hashed} & \vertically{Salted} & \vertically{Hash length} & \vertically{Key stretching} & \vertically{Memory-hard} & \vertically{Random salt} & \vertically{Salt length} \\
\toprule
Hashed & & \stackanchor{\small 36\%}{\small (19)} & \stackanchor{\small 17\%}{\small (9)} & \stackanchor{\small 6\%}{\small (3)} & \stackanchor{\small 2\%}{\small (1)} & \stackanchor{\small 19\%}{\small (10)} & \stackanchor{\small 8\%}{\small (4)} \\
\addlinespace Salted & \stackanchor{\small 100\%}{\small (19)} & & \stackanchor{\small 26\%}{\small (5)} & \stackanchor{\small 11\%}{\small (2)} & \stackanchor{\small 0\%}{\small (0)} & \stackanchor{\small 53\%}{\small (10)} & \stackanchor{\small 21\%}{\small (4)} \\
\addlinespace Hash length & \stackanchor{\small 100\%}{\small (9)} & \stackanchor{\small 56\%}{\small (5)} & & \stackanchor{\small 22\%}{\small (2)} & \stackanchor{\small 0\%}{\small (0)} & \stackanchor{\small 56\%}{\small (5)} & \stackanchor{\small 22\%}{\small (2)} \\
\addlinespace Key stretching & \stackanchor{\small 100\%}{\small (3)} & \stackanchor{\small 67\%}{\small (2)} & \stackanchor{\small 67\%}{\small (2)} & & \stackanchor{\small 0\%}{\small (0)} & \stackanchor{\small 67\%}{\small (2)} & \stackanchor{\small 67\%}{\small (2)} \\
\addlinespace Memory-hard & \stackanchor{\small 100\%}{\small (1)} & \stackanchor{\small 0\%}{\small (0)} & \stackanchor{\small 0\%}{\small (0)} & \stackanchor{\small 0\%}{\small (0)} & & \stackanchor{\small 0\%}{\small (0)} & \stackanchor{\small 0\%}{\small (0)} \\
\addlinespace Random salt & \stackanchor{\small 100\%}{\small (10)} & \stackanchor{\small 100\%}{\small (10)} & \stackanchor{\small 50\%}{\small (5)} & \stackanchor{\small 20\%}{\small (2)} & \stackanchor{\small 0\%}{\small (0)} & & \stackanchor{\small 30\%}{\small (3)} \\
\addlinespace Salt length& \stackanchor{\small 100\%}{\small (4)} & \stackanchor{\small 100\%}{\small (4)} & \stackanchor{\small 50\%}{\small (2)} & \stackanchor{\small 50\%}{\small (2)} & \stackanchor{\small 0\%}{\small (0)} & \stackanchor{\small 75\%}{\small (3)} & \\
\bottomrule
\end{tabular}

  \label{tab:cometric}
\end{table}

\begin{table}\tablestyle
  \caption{Observations of specific hashing methods used by
developers. Some developers recommended multiple hashing methods.}
\begin{tabular}{lr}
\toprule
\multicolumn{1}{c}{Hash} & \multicolumn{1}{c}{Observations}\\
\midrule
Encryption & 9\\
AES & 1\\
MD5 & 6\\
SHA1 & 4\\
SHA256 & 7\\
SHA512 & 1\\
base64 & 3\\
Custom cipher & 1\\
\addlinespace
bcrypt & 6\\
PBKDF2 & 4\\
Argon2 & 1\\
\midrule
Insecure method & 26\\
Secure method & 11\\
\bottomrule
\end{tabular}

  \label{tab:used}
\end{table}

Table~\ref{tab:metric} reports how the teams scored against
Naiakshina's criteria~(Figure~\ref{fig:naiakshina}). In our sample,
only 53 developers (38\%) produced outputs that fulfilled at least one
part of Naiakshina's criteria. The most common criterion fulfilled was
that of hashing data (demonstrated by 38\% of participants who scored
a point, 14\% of overall sample). Just under 20\% of the developers
who scored a point used a random salt or an appropriate hash length
(7\% overall); and the remainder of the points in Naiakshina's
criteria were awarded rarely.

\subsection{RQ1: Do specifications lead to securely stored passwords?}

Developers prompted for a specification ($n=61$) scored better
($\mu=1.03$) than those that were unprompted ($n=77$, $\mu=0.47$)---a
comparison by Mann-Whitney $U$ suggests that
this is a significant difference ($p=0.024$, $U=1947.5$), but with
only a small effect size (rank-biserial
coefficient~\cite{cureton1956rank}, $r_{rb}=0.171$). There remains a
significant difference in performance if we omit the subset of the
unprompted group who reported writing a specification without being asked to---prompted
participants ($n=61$, $\mu=1.03$), in contrast to developers who did
not write a specification ($n=34$, $\mu=0.38$). The two groups are
distinct (Mann-Whitney $U=820$, $p=0.027$) but the effect size remains
small ($r_{rb} = 0.209$). However, a comparison between all
participants who wrote a specification, prompted or not, ($n=104$, $\mu=0.83$) and those
who did not write a specification ($n=34$, $\mu=0.38$) is not statistically significant
($p=0.061$, $U=1495$, $r_{rb}=0.154$).
This could be explained by developers in the \emph{unprompted specification} group (those who were not asked to write a spec but who claimed to have written one anyway) actually writing their spec after their implementation in response to us asking if they had written one beforehand.
This theory is supported by Table~\ref{tab:mwu} where we found no significant difference between
the unprompted specification and the group that claimed not to write a
specification ($p=0.247$).

The distribution of scores is given in
Table~\ref{tab:scores}. 50--70\% of developers did not 
store a password in any meaningfully secure way (a score of 0), and no
developer obtained a perfect score (of 7) using Naiakshina's metric,
although two developers did meet the score indicating best practice (a
score of 6; both were in the \emph{prompted specification} group). Of
the 77 developers whom we did not prompt to write a specification 56\%
(43) claimed to write one anyway unprompted; 26\% (36) of developers
reported referring to some kind of standard or guide when writing
their password storage method; 43\% (59) claimed some formal software
engineering qualification.

\begin{finding} Examining the rank-biserial correlation ($r_{rb}$) to
the scores themselves in
Table~\ref{tab:scores}, suggests that whilst forcing developers to
write a specification before coding will lead to more secure password
storage approaches ($p=0.024$), it isn't going to make a huge
difference---developers \emph{might} remember to hash them or to add
salt, but will still leave them stored insecurely. 
\end{finding}


\subsection{So what else has an effect?}

If the act of forcing developers to write a specification only has a
small impact on their ability to store passwords securely,
then do we find anything else having an effect?

Participants reported their familiarity with cryptography on a 5-point
Likert scale.  There is a small positive relationship between reported
cryptography experience and score (Spearmans
Rho~\cite{spearman1961proof}, $\rho_{s}=0.333$, $p=0.00$), with most
developers reporting that they had \emph{little to no experience}
(107, 78\%) (\emph{No experience}: 44 (32\%), \emph{little
experience}: 63 (46\%), \emph{moderate experience}: 25 (18\%),
\emph{very experienced}: 4 (3\%), \emph{extremely experienced}: 2
(1\%)).  This is in contrast to the findings of Hazhirpasand~et~al{.}
who found no significant relationship between developer experience and
their ability to use a cryptography
API~\cite{hazhirpasand2019impact}---though Hazhirpasand~et~al{.} rated
developer experience on the basis of activity on GitHub, as opposed to
a self-reported value.  We did not find a significant relationship
between developers who had a formal software engineering qualification
and those who did not ($p=0.061$).
Participants
who reported using a standard to inform their code implementation
scored better than those who used no standard but not significantly~($p=0.135$).

\subsection{RQ2: What did developers do?}

Our observations of hashing and salting rates are broadly inline with what
Naiakshina~\etal{} observed~\cite{naiakshina-2017-password}, where an overall
35\% ($\frac{7}{20}$) of developers hashed passwords and 25\% ($\frac{5}{20}$)
also salted them---however Naiakshina~\etal{}'s study explicitly primed half of
their developers ($\frac{10}{20}$) by asking them to store them security, and
only the primed groups hashed or salted their passwords. In contrast, in our
findings we observe similar rates over all participants.

In our study we asked developers to provide code in any programming
language, including pseudocode.  Most developers described their
implementation in these terms using functions called \texttt{hash} and
appending salts, however some gave specific methods for storing their
passwords.  Table~\ref{tab:used} shows the specific methods we
encountered for hashing passwords.  Many developers recommended hash
functions that were inappropriate for password storage\footnote{They are quick to calculate using little memory, thus making them amenable to cracking, unlike memory hard hashes such as PBKDF2.}---including
MD5, the SHA family, and a substitution cipher.  Other developers recommended encryption (which
is unsuitable for password storage~\cite{nist-800-63-3}), or even using base64.  Of all the
developers who stored their passwords hashed, 50\% (26) used an
inappropriate hashing method~\cite{nist-800-63-3}, and only 21\% (11) recommended a secure
modern password hash.  One developer recommended both a secure and insecure method:

\begin{implementation}[1]
\ldots{}hash password in bcrypt or md5\ldots{}
\end{implementation}

\begin{finding} Only a third of developers wrote code to store their
  passwords hashed.  50\% of those developers recommended an insecure
  hash function, and only 21\% recommended a secure hash function. The
  remaining 29\% did not specify the method---they just `hashed' them.
  14\% of developers remembered to salt and hash their passwords.  More
  comprehensive security (Figure~\ref{fig:naiakshina}) was rare.
\end{finding}

\section{RQ3: Why are developers storing passwords like this?}

\begin{table*}\tablestyle
  \caption{Codebook formed from the analysis of developers' explanation of their implementation approach.  Quotes are given to illustrate the use of all
    codes with relevant passages \hltext{underlined}. Some responses were
    assigned more than one code.  No more than 3 codes were used to
    capture any single response.}
  \begin{tabular}{l p{\dimexpr 0.57\linewidth - 2\tabcolsep}r r r}
    \toprule
     \multicolumn{1}{c}{Code}                            & \multicolumn{1}{c}{Description}                                                                                                                   & \multicolumn{1}{c}{Count}                                           & \rotatebox{90}{\stackanchor{Prompted}{Spec}} & \rotatebox{90}{\stackanchor{No}{Spec}} \\  
    \midrule
     Implementation Ease             & The developer wrote it like that as the implementation would be \emph{``simple''}                                             & 36   & 16\% & 29\%\\
    \examplewhy{0}{\hltext{Because it was a simple but quite effective way to store}. To ensure that the data is secure, the function that encrypts the password must be very good.} \\\midrule
     Readability                     & The developer focused on how understandable their code would be to a reader.                                                  & 8           & 7\% & 9\%\\
    \examplewhy{0}{\hltext{I wrote that way because it shows the idea very clearly}. The encryption code is a more difficult question and needs time and ideas to implement a good encryption.} \\\midrule
     Na\"\i{}ve                      & The developer wrote the code in a literal manner without considering the merits of any other approaches.                      & 13                 & 7\% & 12\%\\
    \examplewhy{0}{Because \hltext{I don't know how to wrote code, so just used a literal approach.}} \\\midrule
     Experience                      & The developer made reference to their experience when describing how they wrote their code.                                   & 19            & 11\% & 18\%\\
    \examplewhy{3}{\hltext{I'm somewhat experienced in applied security} and I consider the password should be stored securely, considering the worst case possible.} \\\midrule
     Replication of previous efforts & The developer said they had done it like this before.                                                                           & 8           & 5\% & 6\%\\
    \examplewhy{2}{I wrote code like this \hltext{because it is something I have done before.} I've written a login system for a password manager so recognise that passwords before storage should always be hashed or encrypted to avoid storing them in plain text. I used a struct mainly for storage purposes of this task, but would normally use a database such as SQL to store them, after hashing.}\\
    \midrule
     Feature justification           & The developer justified a specific feature of their implementation (e.g.~ability to send password reminders).                 & 4 & 3\% & 3\%\\
    \examplewhy{0}{This is a simple way to code and \hltext{allow for a reminder to the recipient!}} \\\midrule
     Method justification            & The developer justified the structure of their code.                                                                          & 20  & 18\% & 15\%\\
    \examplewhy{0}{I used a utility class. \hltext{This class stores usernames and passwords in a Map data structure}, and then provides functions for user registration and login.} \\\midrule
     Acknowledgment of limitations  & The developer noted that their code has limitations, and that it doesn't have a certain feature (e.g.~it is insecure).        & 14            & 7\% & 12\%\\
    \examplewhy{3}{It assures the storage of password and allows to recover the password easily \hltext{even if the security is not high.}} \\
    \midrule
     Perceived best practice         & The developer did it this way as this is the correct way to store a password or a standard way in their company.              & 20                   & 18\% & 12\%\\
    \examplewhy{4}{this is \hltext{most accepted way} of storing passwords} \\\midrule
     Consideration of threats        & The developer considered a threat that might attack the code and explicitly attempted to mitigate that threat.                & 18                & 18\% & 9\%\\
    \examplewhy{2}{Hashing passwords is a necessity, \hltext{storing passwords in plain text is a huge security concern:} and should never even be considered.} \\ \midrule
     Taken under advisement          & Someone told them this was a good way to do it.                                                                               & 1            & 0\% & 0\%\\
    \examplewhy{1}{\hltext{my friend who is into cybersecurity told me about this}} \\ \midrule
     Only way I know                 & The developer indicates that this is the only way they knew how to complete the task.                                         & 15       & 11\% & 12\%\\      
    \examplewhy{0}{That was the \hltext{only way i knew} to solve that problem} \\
    \bottomrule                                                                                                                                                                                                                   \\ 
  \end{tabular}
  \label{tab:whycodebook}
\end{table*}

\begin{table*}\tablestyle
  \caption{Codebook from the analysis of developers' responses to what threats did they consider when storing the passwords.  Only developers who indicated that they had considered a threat's responses were analyzed.
  Quotes are given to illustrate the use of all
    codes with relevant passages \hltext{underlined}. Some responses were
    assigned more than one code.  No more than 4 codes were used to
    capture any single response.}
  \begin{tabular}{llrcc}
    \toprule
     \multicolumn{1}{c}{Code} & \multicolumn{1}{c}{Description} & \multicolumn{1}{c}{Count} & \multicolumn{1}{c}{\rotatebox{90}{\stackanchor{Prompted}{Spec}}} & \multicolumn{1}{c}{\rotatebox{90}{\stackanchor{No}{Spec}}} \\ 
    \midrule
     Access & Threat from unauthorized access to the database (e.g. leaks). & 36 & 34\% & 15\%\\
    \examplewhy{6}{\hltext{Password leaks}} \\\midrule
     Cracking & Threat from attacks on stored passwords (e.g. cracking or rainbow tables). & 20 & 16\% & 12\%\\
    \examplewhy{1}{\hltext{reverse the hash code} but i think that is impossible because is unidirectional} \\\midrule
     Hacking & Threat from unspecified threat actors, phishing or social engineering. & 16 & 11\% & 12\%\\
    \examplewhy{0}{Stealing them by a hacker, \hltext{hacked by an unknown user} to steal information and data} \\\midrule
     Programming concerns. & The threat from vulnerabilities in their code (e.g. bugs, SQL injection) & 10 & 8\% & 3\%\\
    \examplewhy{3}{\hltext{SQL injection}, unauthorized DB access} \\\midrule
     Confidentiality & Concerns about making the stored passwords harder to see. & 10 & 10\% & 3\%\\
    \examplewhy{0}{Someone accessing the content that is not the main user. \hltext{If I used strings the password would be stored in strings until} the Garbage Collector clears it and we cannot control when that happens.} \\\midrule
     Malware & Threat from malware, key-loggers or network attacks. & 6 & 5\% & 6\%\\
    \examplewhy{2}{Someone tracing you with \hltext{keylogger or maybe trojan horse}} \\\midrule
     Reflection & Consideration of what they \emph{should} have done and the security of their implementation. & 5 & 3\% & 3\%  \\
    \examplewhy{0}{Since the good is quite simple, \hltext{I am not certain if the storage is secure}.} \\\midrule
     Wider-context & Concerns about the wider impact of an insecurely stored password. & 2 & 2\% & 0\%\\
    \examplewhy{2}{Potential of a database dump, hackers can just login if the passwords were stored in plain text, with hashed passwords they would need to brute force the password. Failure to secure our users passwords \hltext{could lead to them having their accounts on other platforms compromised too} as users tend to reuse passwords.} \\\midrule
     Insider-threat & Threats from insiders who might have access to stored passwords. & 2 & 3\% & 0\%\\
    \examplewhy{4}{The database being accessed by a 3rd party, \hltext{internal threat actors}(excepting those with access to the code for password storage)}\\
    \bottomrule \\
  \end{tabular}
  \label{tab:threatcodebook}
\end{table*}

After implementing their solutions we asked developers why they had
used a particular approach.  Two of the authors used a grounded theory
approach~\cite{Strauss1998} to analyze the responses. Two passes were
required to reach the point of theoretical
saturation~\cite{Glaser1967} when no new codes were identified.  The
resulting codebook, and illustrative examples of each code, is shown
in Table~\ref{tab:whycodebook}.  We also contrasted the distribution
of codes in the \emph{prompted specification} and \emph{no
specification} groups and found them to be broadly similar---with the
\emph{no specification} group being slightly more likely to report
they wrote their code the way they did because the implementation was
easy.  Consequently our remaining
analysis of developers explores why they implemented password storage
in the way they did, and is over the entire study group.

We also asked the developers if 
they considered any threats when implementing their password storage
solution?  Threat modeling is a standard technique when designing for
security~\cite{shostack2014threat} that encourages developers to
consider what defenses are needed to mitigate the potential threats to
a system.  55\% (77)
developers reported considering potential threats when implementing
their password storage solution.  Their responses were analyzed by one
author, again using a grounded theory approach~\cite{Strauss1998}.
Two passes were required to reach the point of theoretical
saturation~\cite{Glaser1967} when no new codes were identified.  The
resulting codebook, again with illustrative examples of each code, is
shown in Table~\ref{tab:threatcodebook}.

\subsection{You either think you do know, or you know you don't know}

Our analysis of the reasons why developers implemented password
storage in the ways they did reveal two interesting
sub-groups. Several answers appear to indicate that
the developers thought they had stored the passwords properly (the
\emph{experience}, \emph{replication of previous efforts},
\emph{perceived best practice} and \emph{taken under advisement}
codes); whereas others seemed to know that their implementation was
limited and that they didn't know how to do it (the \emph{na\"\i{}ve},
\emph{acknowledgment of limitations} and \emph{only way I know}
codes).

Within the group who thought they knew how to store passwords, developers indicated that they believed their approaches were \emph{best practice}.  One developer stored the password directly into a database:
\begin{lstlisting}
INSERT INTO `users` (`username`, `password`) VALUES ('user', PASSWORD('password1'));$\ldots{}$
\end{lstlisting}
They explained this as:

\begin{whyquotation}
  ``Because that is the best way to store the password''
\end{whyquotation}
\noindent
Yet their solution stores the passwords directly without hashing or salting:  they scored 0.
Others stated:

\begin{whyquotation}[4]
  ``this is most accepted way of storing passwords''
\end{whyquotation}
\begin{whyquotation}[0]
  ``It's based on corporate best practices''
\end{whyquotation}

\noindent Developers indicated that they knew their answer was correct because they had done similar tasks before calling upon both their experience as well as previously written code:

\begin{whyquotation}[2]
``Because I wrote a user registration system in the past.''
\end{whyquotation}
\begin{whyquotation}[0]
``I'm used to implementing similar login and authentication mechanisms in university projects and the thought process is always the same:\ldots{}''
\end{whyquotation}
\begin{whyquotation}[2]
``I wrote code like this because it is something I have done
before. I've written a login system for a password manager so
recognize that passwords before storage should always be hashed or
encrypted to avoid storing them in plain text.\ldots{}''
\end{whyquotation}
The relevant part of the code based on the login system for the password manager looked like:
\begin{lstlisting}
user.username = std::cin.get();
user.password = hashPassword(std::cin.get());} $\ldots$
void hashPassword(std::string password) {
 //Cryptography algorithm to hash password, preferably using a salt }
\end{lstlisting}
It hashed the password with a \emph{cryptography algorithm}. It would \emph{preferably} use salt.
Another developer told us that they had taken advice from someone they considered
knowledgeable about cybersecurity:

\begin{whyquotation}[1]
  ``my friend who is into cybersecurity told me about this''
\end{whyquotation}
Yet the solution their friend supplied was mostly inadequate, only
showing signs of hashing (with MD5).
\begin{lstlisting}
$\ldots$ string hashpass = MD5(password);
PasswordDatabase.put("login","password");" 
\end{lstlisting}
This group of developers appear to believe their answers are correct, and that they are following best practice.  They indicate that code similar to what they wrote is in projects they've implemented.  Yet despite this, there are many low scores.  One developer described their experience as:
\begin{whyquotation}[3]
``I have been working as a software engineer for 8 years and have developed authentication systems for our clients hundreds of times so have come to learn the best practices for doing so.''
\end{whyquotation}
Their score would suggest they have more to learn.

Not all developers seemed to be so confident.
In contrast to the first group, other developers gave explanations that suggest they are aware that they don't know how to store passwords properly, or at least that their code had limitations.
For example one developer stated:
\begin{whyquotation}[1]
``I don't have a lot of knowledge about password storaging\ldots''
\end{whyquotation}
Their implementation hashed the password, suggesting the developer was confused about the distinction between \emph{hashing} and \emph{encryption}:
\begin{lstlisting}
$...$ string encpass = anHashingFunction(password);
myfile << username << endl;
myfile << encpass << endl;
\end{lstlisting}
They scored 1 point using Naiakshina's criteria, for hashing the password.
Another said:

\begin{whyquotation}[1]
``I wrote the code that way since it's the only way I know how to check if the passwords are valid, and the hashing /~storing bit because unhashed passwords are unsafe.
%
\ldots{}
Other methods could be used to encrypt the password, but I've heard hashing or MD5 hashing is the most common.'' 
\end{whyquotation}
The following two explanations came from developers who were in the top 10\% of highest scoring implementations, according to Naiakshina's criteria.  Both acknowledge the security of their implementations, and that it wasn't perfect:
\begin{whyquotation}[3]
``It assures the storage of password and allows to recover the password easily even if the security is not high.''
\end{whyquotation}
\begin{whyquotation}[4]
  ``It was the simplest and easiest way I could think
  \ldots{}
  This way it protects most cases, but of course a more elaborate with more defence lines is needed (and the salt implementations is not very well done, \ldots{}
  )''
\end{whyquotation}
This group of developers form a counterpoint to the first group who
think they know how to store passwords: they know they don't know
everything.  Whilst directly comparing groups is hard (the codes are
not independent, and emerge from what developers said) a comparison of
average score between them suggests that neither group is storing
passwords more securely (comparison of mean score between the
\emph{think they know} and \emph{know they don't know} groups: $0.79$
vs $0.63$). In short, roughly a third of developers appear to be
overly confident in their knowledge of best practice in our study.
Despite this their answers do indicate that developers are aware that
password storage is an inherently security oriented task.  They know they should be storing passwords securely, but plenty of them are overconfident and have misplaced assurance in what they do.

\subsection{On reflection, perhaps you know}

The \emph{acknowledgment of limitations} code from Table~\ref{tab:whycodebook} and the \emph{reflection} code from Table~\ref{tab:threatcodebook} are interesting as they highlight when developers indicated that their implementation was lacking security aspects.
For example, one developer
remembered to use a hash function with a suitable length in their
implementation.  When stating which threats they considered they note
that they had forgotten to salt the password (and why that was
necessary).

\begin{whyquotation}[2]
  ``\ldots
  (Thinking about it,
 it might have been a good idea to concatenate some constant text at
 the end of the password so that whether the user uses the same
 password on two different attacked services cannot be determined
 simply by checking whether the hashes are identical.)''
\end{whyquotation}

Others, on reflection, realized their
solution was inadequate (with respect to security).  Both of the
following developers stored their passwords directly as plain-text
(scoring 0), yet when asked to consider the threats they later seemed to
realize that there were some they should have considered:

\begin{whyquotation}[0]``Unfortunately, I have not considered any
threats, but I know that the password should be
encrypted.''\end{whyquotation}

\begin{whyquotation}[0]``Actually I didn't consider them in the
pseudo-code but I assume there are some threats like brute
hacking''\end{whyquotation}

\noindent One developer stored their password directly,
but when considering threats gave a guide to storing them that would
have scored at least 3:

\begin{whyquotation}[0]``The best security practice is not to store
the password at all (not even encrypted), but to store the salted hash
(with a unique salt per password) of the encrypted
password.''\end{whyquotation}

Whilst we did not prime developers for storing passwords securely as
Naiakshina did~\cite{naiakshina-2019-if-you-want}, we still found that
developers talked about the security of their implementation.  Around
half of the developers reported considering threats when implementing
their solutions, and some made reference to those threats when
describing why they'd implemented the code in the way that they did.
The threats described in Table~\ref{tab:threatcodebook} are
reasonable: the reason we hash and salt passwords is to ensure if the
database is \emph{accessed} illegally, that the passwords cannot be
trivially \emph{cracked}. This is hopeful: it suggests developers may
be learning that passwords and security are linked and that they
should store them securely, not \emph{if we want} them
to~\cite{naiakshina-2019-if-you-want}.  Developers may not realize it
immediately that passwords should be stored securely, but if given
time to reflect they do seem to make that connection.  Even if a
developer doesn't initially realize that password storage is a
security oriented task, by giving them time to reflect (in complement
with time to consider a specification) some developers do realize that
passwords must be stored securely.

\subsection{Google is your friend}

\begin{table}\tablestyle
  \caption{Sites apparently referenced by developers.}
  \begin{tabular}{lr}
\toprule
\multicolumn{1}{c}{Source} & \multicolumn{1}{c}{Count} \\
\midrule
\texttt{stackoverflow.com} & 12\\
\texttt{gist.github.com} & 1\\
\texttt{docs.microsoft.com} & 1\\
\texttt{simplecode121.blogspot.com} & 1\\
\texttt{www.programcreek.com} & 1\\
\texttt{happycoding.io} & 1\\
\texttt{www.baeldung.com} & 1\\
\texttt{pypi.org} & 1\\
\texttt{www.codota.com} & 1\\
\texttt{medium.com} & 1\\
\texttt{www.the-art-of-web.com} & 1\\
\texttt{www.tutorialspoint.com} & 1\\
\texttt{www.cpp.re} & 1\\
\texttt{docs.python.org} & 1\\
\texttt{www.w3resource.com} & 1\\
\texttt{howtodoinjava.com} & 1\\
\bottomrule
\end{tabular}

  \label{tab:sources}
\end{table}

There is much guidance and advice online about how to store passwords
(on sites such as \emph{Stack Overflow}, for example).  When analyzing developers reasons and implementations, we checked for copying from such sites.
We searched online to see if any of the
implementations and pseudocode developers provided appeared online and
found 12 appearing on the \emph{Stack Overflow} developer forum
alongside 15 others appearing on other websites
(Table~\ref{tab:sources}).  We found that 27 developers (20\%)
appeared to have copied code from various sites (shown in
Table~\ref{tab:sources}), with the majority having taken code directly
from Stack Overflow. Of these 27, 7 reported using a standard.


The group that used the online source
appeared to score significantly higher than the group whose source was unknown ($\mu=1.19$ vs $\mu=0.60$,
$p=0.017$) though the effect size was relatively small ($U=1155$,
$r_{rb}=0.23$); however when reading the solutions online we noticed that some of the articles developers appeared to have copied from also contained guidance on how to store them near-perfectly (according to Naiakshina's criteria).
One developer justified their answer as:

\begin{whyquotation}[2]
  ``I wrote the code like this because it is good practice not to store a password in clear.''
\end{whyquotation}
The solution appeared to have been taken from a
\emph{how-to} site which described how to implement password storage
with a variety of hashes and salts, starting with MD5 and ending with
bcrypt and scrypt; however the site went on to describe a solution at
the end of the article that would have scored 6~points (losing the
last point for only using 16 instead of 32~bits for the salt).

Another developer described their implementation as:
\begin{whyquotation}[0]
  ``The first and foremost way to store passwords in your database is to
  have the plain text.\ldots''
\end{whyquotation}
This came from a blog
post~\cite{cox2017password}. The remainder states:

\begin{whyquotation}
  ``(don't do this)
  I can't emphasize strongly enough that you should NEVER, EVER, store passwords in plain text."
\end{whyquotation}
The article does describe how to store
passwords using a hash and a randomly generated salt (3 points); yet again the developer only copied the
insecure counter-example at the start of the article.%
If we want
developers to store passwords correctly then we need to make sure the
code we want them to copy is immediately obvious.
That
developers are copying code from online isn't of itself
worrying---if they copied \emph{the right} solutions we
might see secure password storage.  Instead, some developers seem to
be copying online code, using the articles to justify themselves, but
not reading the article all the way through. 

\section{Discussion}
\subsection{Do specifications lead to securely stored passwords?}

Writing a specification has a small, but positive, effect on
developers ability to store passwords securely.  Yet in saying this,
we avoid the bigger issue that developers seem to really struggle with
implementing password storage correctly.  In our study 62\% of
developers failed to hash, salt or add any security mechanism
whatsoever.

Our paper joins an ever-growing body of work demonstrating that
developers are struggling implementing password
storage~\cite{wijayarathna2018johnny,naiakshina-2017-password,naiakshina-2019-if-you-want};
but our work also finds that it isn't \emph{just} that developers
struggle to use cryptography
APIs~\cite{green-2016-developers,patnaik2019usability}; and it isn't
\emph{just} that developers don't know enough about cryptography to
complete the task correctly:  developers stated that hashing passwords
with MD5 was best practice (it isn't~\cite{nist-800-63-3}).
Developers would forget about salting and say that's the way that they do
it in their company.  They would claim cybersecurity expertise, to
have password storage code in production, as reasons why
their code is secure; as reasons why their code follows best
practice---and yet they fall short.  Our paper finds that specification is beneficial ($p=0.024$),
but, equally importantly, it highlights that developers don't know that they're storing
passwords insecurely.

\subsection{Beyond Naiakshina's criteria}

In this study we measured developers' ability to store passwords using
Naiakshina's criteria, as a proxy for the NIST~SP~800-63-3
standard~\cite{nist-800-63-3} which defines current best practice.  The criteria
and standard itself is somewhat quiz-like, asking developers to remember
cryptographic techniques like hashing and salting as well as arbitrary lengths and
counts. If developers do not know these requirements then they will not
remember them. so what then do we learn \emph{beyond} the fact that developers do not
seem to recall Naiakshina's criteria?
An ideal specification might have listed the criteria in full as functional
requirements, but it might also have been as simple as:
\emph{``store the password securely, following NIST SP 800-63-3.''};
yet none of the developers in our
study made reference to any standard in their specifications.  Developers seemed
to know there was a \emph{best practice} they ought to be following, yet didn't
appear to go look up what it actually was.  Whether this
generalizes and developers' recollection of other standards is equally poor is a
topic for future work.
Whilst one good approach to implementing password storage is to do what the
standard says, another equally valid (and arguably better) aproach is to use a
framework and let it do it for you.  Web-frameworks, like Django, include
password storage systems (and in Django's case explicitly reference NIST
password
standards~\cite{django-password-management})
and can take care of passwords for developers.  Again, developers in our study
did not appear to make use of frameworks like this, so is it that developers are
unaware of these features inside frameworks, or did they choose not to use them?
Perhaps given the seeming recalcitrance towards reading standards, the
resistance towards using frameworks, and the confidence many displayed that they
were in fact following best practice we might conclude that developers are over
confident in their abilities.  Why use a library when you can implement it
yourself trivially?  Why check the standard when you know already what best
practice is?  Developers seem to have learned not to \emph{roll their own
crypto}; perhaps they should also consider avoiding \emph{rolling their own
authentication} in future too?

\subsection{Developers are still not the enemy}

We say that \emph{users and developers are not the
enemy}~\cite{adams1999users,green-2016-developers}---that we must not
blame users or developers when an API or security interface is not
designed for a human to be able to use correctly.  Yet when we talk
about password storage we present it as a list diktats that developers
\emph{must} implement to ensure they do the task correctly.  As we,
and
others~\cite{wijayarathna2018johnny,naiakshina-2017-password,naiakshina-2019-if-you-want},
have shown developers cannot follow these instructions.  Perhaps then,
instead of pointing out that developers can't store passwords and
providing lists on what they \emph{must} do, we should \emph{fit
the task to the human} and provide alternative mechanisms for storing
passwords correctly without having to remember what the current best
practice actually is, or understand the intricacies of various hashing
schemes. Truman said of being the President: \emph{``He'll sit here
and he'll say, ``Do this! Do that!'' And nothing will
happen''}~\cite{neustadt1960presidential}; and a comparison can be
drawn to the security and cryptography communities: we cannot keep
sitting here; saying, \emph{``Hash this! Salt That!''} and pointing at
NIST SP 800-63-3, because Truman was right: \emph{``nothing will
happen''}.  We need to find usable mechanisms for password storage.
What might these mechanisms look like? Cryptography libraries, such
as Google's \emph{Tink}~\cite{google-tink}, are attempting to wrap
cryptographic details so that developers can use cryptography without
understanding what a hash really is~\cite{schmieg2020tweet}. There has
been limited usability validation of such
approaches~\cite{mindermann2020fluid}, however, and further work,
documentation, and exemplar code~\cite{patnaik2019usability} is needed
to show whether this approach is effective.  Alternatively, some
developers seemed to copy code from online sources---ensuring that
developers can find the trivially find the \emph{right} way to store
passwords and that the \emph{right} code is trivially available may
also help developers without requiring them to understand the
cryptographic details.
Finally, a different solution altogether may be to encourage
developers not to store passwords at all and instead use federated
identity management systems (such as OAuth~\cite{hardt2012oauth}).
Whilst these systems can remove the need for some app developers to
implement cryptography correctly, they come with their own set of
privacy and security \emph{gotchas}~\cite{lodderstedt2020oauth} and
challenges~\cite{sun2012devil}---we should be cautious that by
recommending an alternative to passwords we are not replacing the
challenge of storing a password with the challenge of implementing a
federated authentication system.  Work on privacy-preserving federated
identity management has helped to resolve some of the privacy
challenges associated with federated identity
management~\cite{chow2012spice,isaakidis2016unlimitid}, though these
are yet to be widely adopted in practice.

Explicitly prompting developers to write a specification does help
improve the quality of password storage; but developers are still
mostly failing at password storage whilst still believing they are
getting it right. Giving developers time to reflect helps them realize
the limitations of their approach: but until we have
\emph{developer-centered usable} password storage methods, the problem
of poorly stored passwords isn't going away. We can do better than saying \emph{``Do this! Do that!''} and watching nothing happen.

\section{Conclusion}

Does the act of writing a specification (how the code should behave) for a piece
of security sensitive code lead to developers producing more secure code?  In a
statistical sense: yes, though the effect is small ($p=0.027$, $r_{rb}=0.209$).
In a broader sense however we show that whilst writing a specification does help
developers remember more of the conditions for secure password storage, leaving
this task to a memory exercise and hoping developers refer to a standard isn't
working.
Future work should examine and empirically evaluate alternative strategies for helping developers complete authentication tasks---whether in the form of usable cryptography libraries, privacy preserving federated identity schemes, or alternative awareness schemes to diktats and standards.
Additionally, whilst this study looked to see if \emph{any} form of specification improved developers ability to store passwords correctly; \emph{specific} approaches (whether that be software building codes~\cite{landwehr2015we-need}, requirements engineering, or formal verification) may yield more promising results. Finally, in this study we saw developers struggling to remember how to do secure password storage, but we may see similar results for other areas where knowledge of what the \emph{right thing to do} is conveyed only through diktats and standards---future work should examine whether this result is general or specific to password storage.

\section*{Acknowledgment}
This research is supported in part by
EPSRC Grant
EP/P011799/2
and the National Cyber Security Centre.

\balance
\bibliographystyle{IEEEtran}
\bibliography{bibliography}
\end{document}